\documentclass[twocolumn,showpacs,floatfix,superscriptaddress,amsmath,amssymb,prl]{revtex4}
\usepackage{amsfonts}
\usepackage{amsmath}
\usepackage{txfonts}
\usepackage{amssymb}
\usepackage{graphicx}
\usepackage{hyperref}
\usepackage{overpic}
\usepackage{ulem}

\newcommand{\be}{\begin{equation}}
\newcommand{\ee}{\end{equation}}

\begin{document}

\title{Singularities in ground state fidelity and quantum phase transitions for the Kitaev model}

\author{Jian-Hui Zhao}
\affiliation{Centre for Modern Physics and Department of Physics,
Chongqing University, Chongqing 400044, The People's Republic of
China}

\author{Huan-Qiang Zhou}
\affiliation{Centre for Modern Physics and Department of Physics,
Chongqing University, Chongqing 400044, The People's Republic of
China}

\begin{abstract}
The ground state fidelity per lattice site is shown to be able to
detect quantum phase transitions for the Kitaev model on the
honeycomb lattice, a prototypical example of quantum lattice systems
with topological order. It is found that, in the thermodynamic
limit, the ground state fidelity per lattice site is non-analytic at
the phase boundaries: the second-order derivative of its logarithmic
function with respect to a control parameter describing the
interaction between neighboring spins is logarithmically divergent.
A finite size scaling analysis is performed, which allows us to
extract the correlation length critical exponent from the scaling
behaviors of the ground state fidelity per lattice site.
\end{abstract}

\date{\today}

\maketitle

With the advent of its discovery in the fractional quantum Hall
effect~\cite{tsui}, topological order emerges as a new paradigm in
the study of quantum phase transitions (QPTs)~\cite{wen}. Subsequent
investigations show that topological order occurs in various
strongly correlated lattice systems undergoing QPTs. A
characteristic feature of quantum systems with topological order is
their insensitivity to any \textit{local}
perturbations~\cite{ortiz}. Such an essential difference between
topological and symmetry-breaking orders invalidates  the usual
tools used to describe a symmetry-breaking order, such as long range
correlations, broken symmetries, and local order
parameters~\cite{mapping}.

Recently, much attention has been paid to an exactly solvable spin
1/2 model on a honeycomb lattice introduced by Kitaev~\cite{kitaev}
for fault-tolerant topological quantum computation~\cite{tqc}. The
model describes a set of spins located at the vertices of a
two-dimensional honeycomb lattice, subject to a spatially
anisotropic interaction between neighboring spins. It has been shown
that it carries excitations with both Abelian and non-Abelian
braiding statistics, which do not obey ordinary bosonic and
fermionic statistics, but are anyons with more intricate statistical
behavior~\cite{anyon}. An experimentally feasible realization of the
model in a system of cold atoms on an optical lattice has been
addressed~\cite{dassarma} (see also~\cite{duan,zoller}), with the
expectation to perform quantum computation by utilizing braiding of
collective excitations implanted in topologically ordered coherent
quantum many-body states.

On the other hand,  a viable scheme to determine the ground state
phase diagram of a quantum lattice system without prior knowledge of
order parameters was proposed in
Refs.~\cite{zhou,zov,zzwl,orderpara}. This was achieved by studying
the singularities in the ground state fidelity per lattice
site~\cite{numerical}. In fact, the ground state fidelity may be
interpreted as the partition function of a classical statistical
vertex model with the same lattice geometry by using the tensor
network representations of quantum many-body wave
functions~\cite{zov}. Therefore, the fidelity per lattice site is
nothing but the partition function per site in the classical
statistical vertex lattice model~\cite{baxter}. This justifies why
QPTs may be detected as singularities in the fidelity per lattice
site as a function of the control parameters (see also
Refs.~\cite{fidelity,fidelity1} for the connection between the
fidelity and QPTs). Therefore, an intriguing question is to see if
the fidelity approach captures the physics underlying QPTs in
quantum lattice systems with topological order.

The purpose of this paper is to show that the ground state fidelity
per lattice site is able to detect QPTs for the Kitaev model on the
honeycomb lattice, a prototypical example of quantum lattice systems
with topological order. First, we derive the ground state fidelity
per lattice site between different ground states from the exact
solution of the Kitaev model on the honeycomb lattice. This is
achieved by exploiting the fact that the original spin model on the
honeycomb lattice is rephrased as a $p$-wave BCS model with a
site-dependent chemical potential for spinless fermions on a square
lattice~\cite{chen} (see also Refs.~\cite{xiang,hu,pachos}). The
ground state of the latter is a BCS type state, as a consequence of
the Jordan-Wigner, Fourier and Bogoliubov transformations. Second,
the phase boundaries separating the gapless phase from different
gapful phases are reproduced by investigating the singularities in
the fidelity per lattice site as a function of the control
parameters. It is found that, in the thermodynamic limit, the ground
state fidelity per site is non-analytic at the phase boundaries.
That is, the second-order derivative of its logarithmic function
with respect to a given control parameter is logarithmically
divergent as the phase boundaries are crossed. Third, we perform a
finite size scaling analysis for the Kitaev model, aiming at
extracting the correlation length critical exponent from the scaling
behaviors of the ground state fidelity per site. Our exact results
offer a benchmark to investigate QPTs for two-dimensional quantum
lattice systems with topological order numerically in the context of
tensor network representations~\cite{iTEBD,iPEPS,mera1,mera2}.

{\it The Kitaev model on a honeycomb lattice.} Consider a spin 1/2
model on a honeycomb lattice with the Hamiltonian~\cite{kitaev}
\begin{eqnarray}
H &=& -J_x \sum_{x-bonds}  \sigma^x_i \sigma^x_j -J_y \sum_{y-bonds}
\sigma^y_i \sigma^x_j \nonumber\\
&& -J_z \sum_{z-bonds}  \sigma^z_i \sigma^z_j, \label{kitaev}
\end{eqnarray}
where $J_\alpha$ are interaction (control) parameters and
$\sigma^\alpha_j$ are the Pauli matrices at the site $j$, with
$\alpha = x,y$ and $z$. The Hamiltonian (\ref{kitaev}) may be
fermionized by performing the Jordan-Wigner
transformation~\cite{xiang,hu,pachos,chen} from the Pauli spin
matrices $\sigma^\alpha_j$ to the spinless fermion operators
$c_j^\dagger$ and $c_j$. This one-dimensional fermionization is
realized by deforming the hexagonal lattice into a brick-wall
lattice which is topologically equivalent to the original lattice.
If we introduce the Majorana fermions: $A=(c-c^\dagger)/i$ and
$A=c+c^\dagger$, then the Hamiltonian (\ref{kitaev}) becomes
\begin{eqnarray}
H&=& -iJ_x \sum_{x-bonds}  A_w  A_b  +i J_y \sum_{y-bonds}  A_b  A_w
\nonumber\\
&&  -i J_z \sum_{z-bonds}  \alpha_r A_b  A_w, \label{majorana}
\end{eqnarray}
where the subscripts $w$ and $b$ denote two sublattices in the
brick-wall lattice, and $\alpha_r \equiv B_bB_w$ along the $z$-bond
is conserved~\cite{xiang}, with $r$ being the coordinate of the
midpoint of the bond connecting the $b$-type and $w$-type sites.
This in turn is equivalent to a model of spinless fermions on a
square lattice with a site-dependent chemical potential:
\begin{eqnarray}
H &=& J_x \sum_{r}  (d^\dagger_r + d_r) (d^\dagger_{r+\hat{e}_x}
-d^\dagger_{r+\hat{e}_x})\nonumber\\
&& + J_y \sum_{r}  (d^\dagger_r + d_r)
(d^\dagger_{r+\hat{e}_y}-d^\dagger_{r+\hat{e}_y})\nonumber\\
&&  + J_z \sum_r \alpha_r (2 d^\dagger d_r -1). \label{fermion}
\end{eqnarray}
Here the unit vector $\hat{e}_x$ and $\hat{e}_y$ connects two $z$
bonds and crosses a $x$- and $y$-bond, respectively.  For large
enough systems, the ground state configurations are bulk vortex-free
configurations~\cite{kitaev,pachos}, which implies $\alpha_r =1$ for
all $r$. Therefore, the ground state may be obtained by performing a
fourier transformation. Up to an unimportant additive constant, the
Hamiltonian (\ref{fermion}) in the vortex-free configuration now
reads,
\begin{equation}
 H_g = \sum _k \left( \epsilon_k d^\dagger d_k + i \frac {\Delta_k}{2}
 \left( d^\dagger_k d^\dagger_{-k } + \rm {H.C.} \right) \right),\label{BCS}
\end{equation}
with $\epsilon_k = 2 J_z - 2J_x \cos k_x - 2J_y \cos k_y$, and
$\Delta_k = 2J_x \sin q_x + 2J_y \sin k_y$. The Hamiltonian
(\ref{BCS}) is a $p$-wave type BCS pairing model and can be
diagonalized by means of the Bogoliubov transformation. It yields
that the BCS type ground state is  $|g \rangle = \prod_k (u_k +v_k
d^\dagger_k d^\dagger_{-k}) |0\rangle$, where $|u_k|^2=1/2
(1+\epsilon_k/E_k)$ and $|v_k|^2=1/2 (1-\epsilon_k/E_k)$, with the
quasiparticle excitation energy $E_k = \sqrt
{\epsilon_k^2+\Delta^2_k}$~\cite{chen}.

{\it The ground state fidelity per lattice site.} Consider two
ground states $|g\rangle$ and $|g'\rangle$ corresponding to
different values of the control parameters $\vec{J} \equiv
(J_x,J_y,J_z)$ and $\vec{J}' \equiv (J'_x,J'_y,J'_z)$, respectively.
The fidelity $F(\vec{J};\vec{J}') \equiv \langle g'|g\rangle$
asymptotically scales as $F(\vec{J};\vec{J}') \sim
{d(\vec{J};\vec{J}')}^N$, with $N$ the total number of sites in the
lattice. Here $d(\vec{J};\vec{J}')$ is the ground state fidelity per
lattice site, introduced in Refs.~\cite{zhou,zov}. Although
$F(\vec{J};\vec{J}')$ becomes trivially zero for continuous QPTs,
the fidelity per lattice site is well defined in the thermodynamic
limit:
\begin{equation}
d(\vec{J};\vec{J}') = \lim_{N \rightarrow \infty}
F^{\frac{1}{N}}(\vec{J};\vec{J}'). \label{d}
\end{equation}
It satisfies the properties inherited from the fidelity
$F(\vec{J};\vec{J}')$: (i) normalization $d(\vec{J};\vec{J})=1$;
(ii) symmetry $d(\vec{J};\vec{J}')=d(\vec{J}';\vec{J}$; and (iii)
range $0 \le d(\vec{J};\vec{J}')\le 1$.

For the Kitaev model on the honeycomb lattice, the logarithmic
function of the fidelity per site, $\ln d_h(\vec{J};\vec{J}')$, is
\textit{half} of the logarithmic function of the fidelity per site,
$\ln d_{sq}(\vec{J};\vec{J}')$, for the model of spinless fermions
on a square lattice. This results from the fact that the number of
sites in the honeycomb lattice doubles that of sites in the square
lattice. The BCS type ground state $|g \rangle$ yields the ground
state fidelity per lattice site for the spinless fermion model on
the square lattice:
\begin{equation}
\ln d_{sq}(\vec{J};\vec{J}') = \frac {1}{(2\pi)^2} \int^\pi_0 dk_x
\int^\pi_0 dk_y \ln (u_k^* u'_k +v_k^* v'_k),
 \label{kitaev-d}
\end{equation}
where $u_k$ and $ v_k$ depend on $\vec{J}$, whereas $u'_k$ and $
v'_k$ depend on $\vec{J}'$. Here we emphasize that although the
information about the topological nature of the Kitaev model is lost
in the spinless fermion representation, the unitary equivalence
between the two representations preserves the fidelity. Since the
extra prefactor does not affect the singularities in $\ln
d_h(\vec{J};\vec{J}')$ and $\ln d_{sq}(\vec{J};\vec{J}')$, hereafter
we focus on $\ln d_{sq}(\vec{J};\vec{J}')$ to carry out the scaling
analysis below, and  omit the subscripts for brevity.

For a finite-size system, the Hamiltonian (\ref{BCS}), resulted from
the Jordan-Wigner, Fourier and Bogoliubov transformations, depends
on boundary conditions imposed on the original spin model
(\ref{kitaev}). In contrast to open boundary conditions, there is an
extra boundary term if one adopts the periodic boundary conditions.
However, such a boundary term does not contribute to the fidelity
per site, although it carries the topological dependence of the
ground state degeneracy~\cite{chen}. From now on, we are only
concerned with the fermion model on a square lattice with the
periodic boundary conditions (i.e., a torus) to analyze the ground
state fidelity per lattice site for finite-size systems~\cite{mera},
from which it is sufficient to extract the bulk behaviors of the
model. As such, for a system on a torus with an even linear size
$L$, the logarithmic function of the ground state fidelity per
lattice site, $\ln d(\vec{J};\vec{J}')$, takes the form:
\begin{equation}
\ln d(\vec{J};\vec{J}') = \frac {1}{L^2} \sum _{k_x,k_y} \ln (u_k^*
u'_k +v_k^* v'_k).
 \label{kitaev-d-finite}
\end{equation}
Here $k_x$ and $k_y$ take values from the set: $\pi m/L$$
(m=-(L-1)/2,\dots,(L-1)/2)$, and the double summation is over all
positive values of both $k_x$ and $k_y$.

\begin{figure}[ht]
\begin{overpic}[width=42mm,totalheight=36mm]{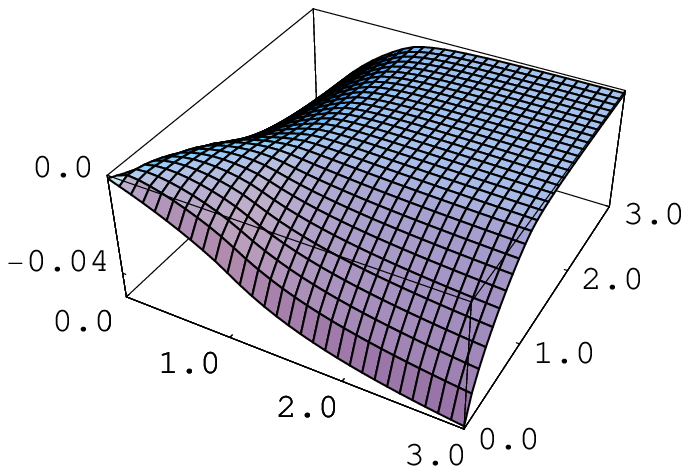}
\put(0,65){$(a)$} \put(2,46){$\ln d$} \put(34,11){$J_x$}
\put(92,22){$J_x'$}
\end{overpic}
\hspace{0in}
\begin{overpic}[width=42mm,totalheight=36mm]{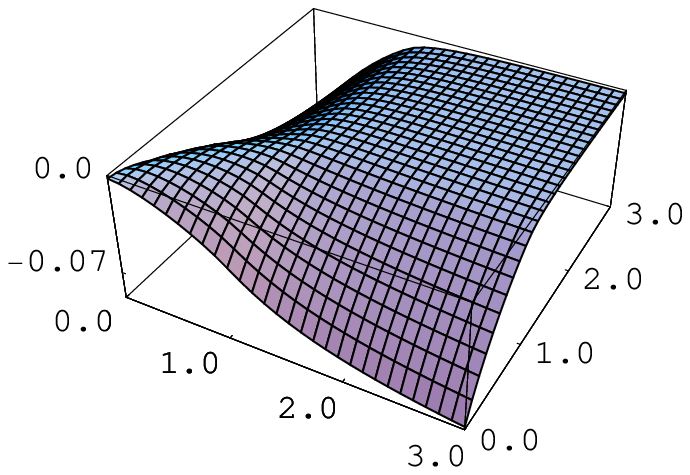}
\put(0,65){$(b)$} \put(2,46){$\ln d$} \put(34,11){$J_z$}
\put(92,22){$J_z'$}
\end{overpic}
\caption{(color online) (a) The logarithm of the fidelity per
lattice site, $\ln d(\vec{J},\vec{J}')$, is shown as a function of
$J_x$ and $J'_x$ for fixed $J_y=J'_y=J_z=J'_z=1/2$. It exhibits a
pinch point at $(J_{xc},J_{xc})=(1,1)$. (b) The logarithm of the
fidelity per lattice site, $\ln d(\vec{J},\vec{J}')$, is shown as a
function of $J_z$ and $J'_z$ for fixed $J_x=J'_x=J_y=J'_y=1/2$. It
exhibits a pinch point at $(J_{zc},J_{zc})=(1,1)$. Here a pinch
point is defined as an intersection of two singular
lines.}\label{fig1}
\end{figure}

{\it Ground state phase diagram and singularities in the ground
state fidelity per lattice site.} Now we turn to the ground state
phase diagram. This follows from the singularities in $\ln
d(\vec{J};\vec{J}')$. One may show that $\ln d(\vec{J};\vec{J}')$ in
Eq. (\ref{kitaev-d}) and the first-order derivative with respect to
a control parameter is continuous, but the second-order derivative
logarithmically diverges when the phase boundaries determined by
$|J_x| = |J_y|+|J_z|, |J_y| = |J_z|+|J_x|$ and $ |J_z| =
|J_x|+|J_y|$ are crossed.  This is consistent with the original
analysis by Kitaev~\cite{kitaev} (see also
Refs.~\cite{chen,xiang,pachos}). In Fig.~\ref{fig1}(a), we plot the
logarithm of the fidelity per site, $\ln d(\vec{J};\vec{J}')$, as a
function of $J_x$ and $J_x'$ for $J_y=J_z=1/2$ and $J_y'=J_z'=1/2$.
It is seen that a pinch point occurs at $(J_{xc},J_{xc})=(1,1)$.
That is, there are singularities along the lines $J_x=1$ and
$J'_x=1$. Therefore, the drastic change of the ground state
many-body wave functions at $J_{xc}$ is reflected as the
singularities in $\ln d(\vec{J};\vec{J}')$. Similarly, the numerical
results are plotted in Fig.~\ref{fig1}(b) for the logarithm of the
fidelity per lattice site, $\ln d(\vec{J},\vec{J}')$, as a function
of $J_z$ and $J'_z$ for fixed $J_x=J'_x=J_y=J'_y=1/2$, with a pinch
point at $(J_{zc},J_{zc})=(1,1)$.

\begin{figure}[ht]
\begin{overpic}[width=38mm,totalheight=28mm]{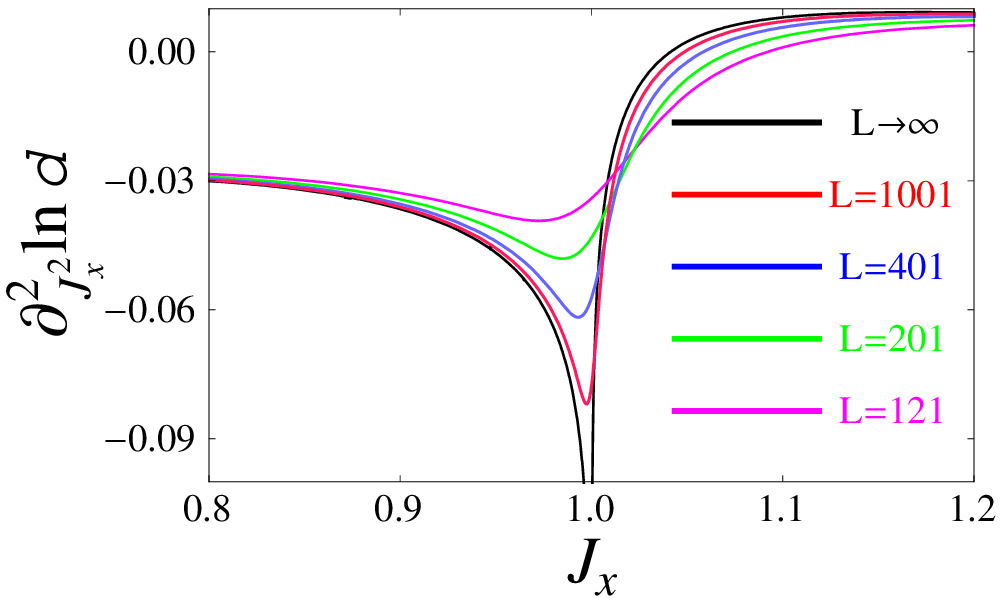}
\put(0,70){$(a)$}
\end{overpic}
\hspace{0in}
\begin{overpic}[width=38mm,totalheight=28mm]{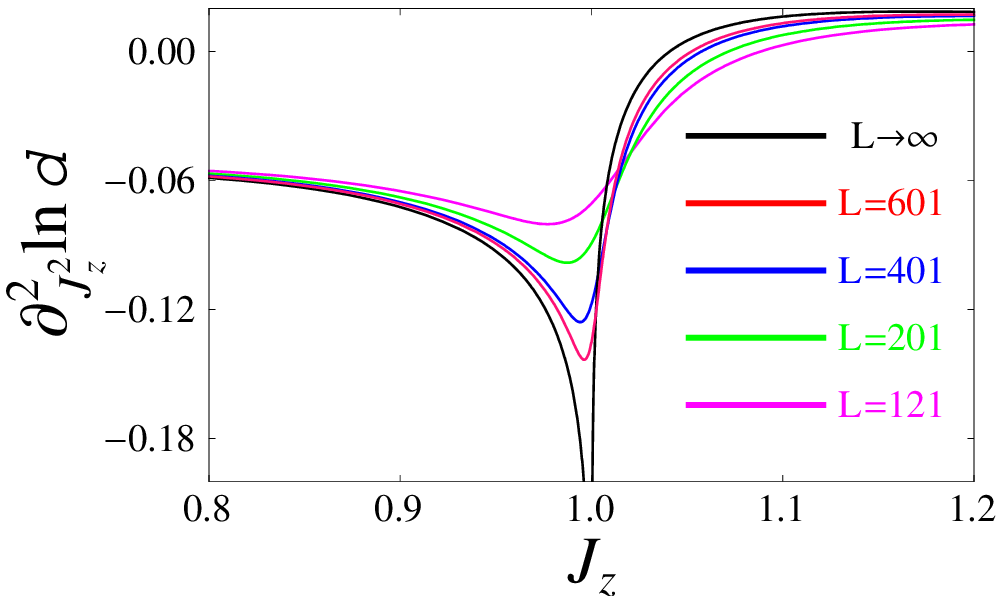}
\put(0,70){$(b)$}
\end{overpic}
\hspace{0in}
\begin{overpic}[width=38mm,totalheight=28mm]{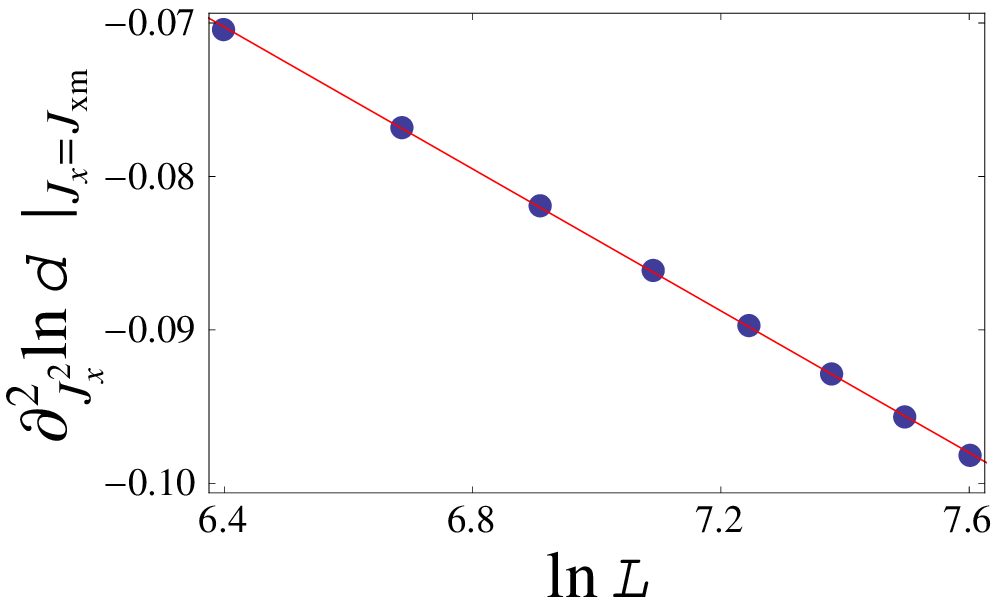}
\put(0,70){$(c)$}
\end{overpic}
\begin{overpic}[width=38mm,totalheight=28mm]{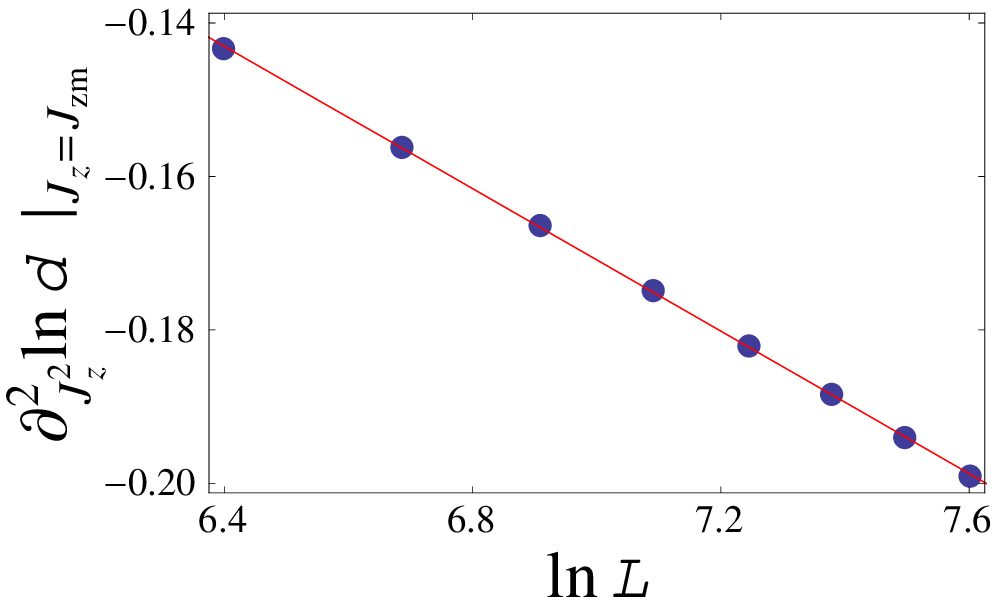}
\put(0,70){$(d)$}
\end{overpic}
\hspace{0in}
\begin{overpic}[width=38mm,totalheight=28mm]{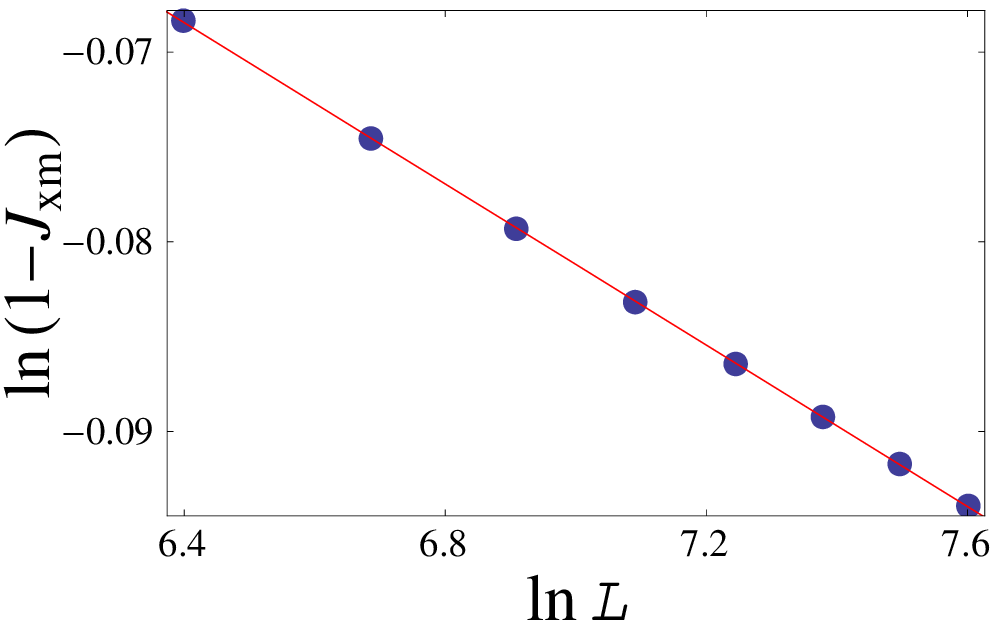}
\put(0,70){$(e)$}
\end{overpic}
\hspace{0in}
\begin{overpic}[width=38mm,totalheight=28mm]{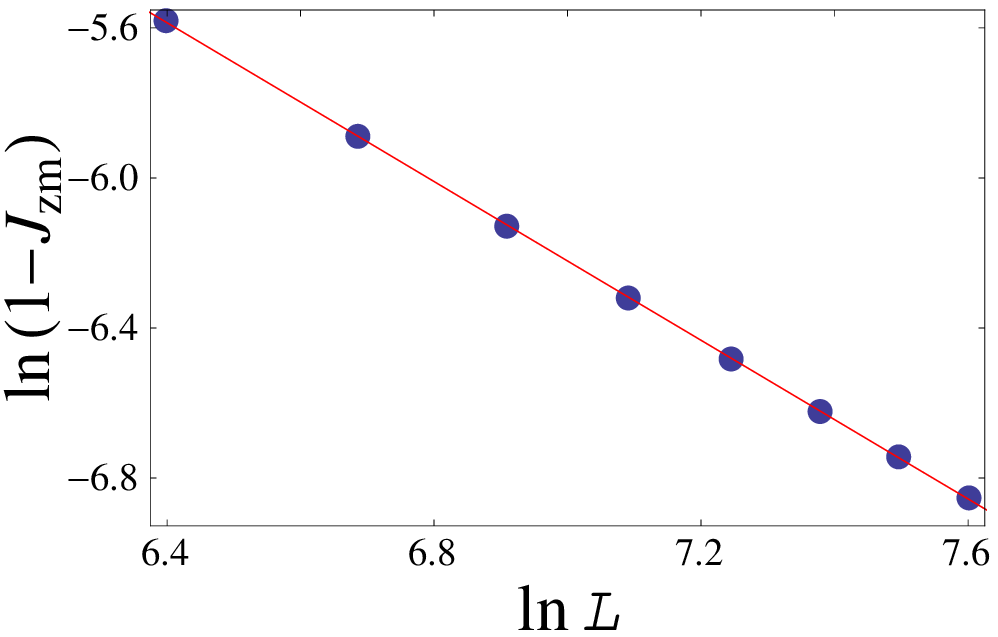}
\put(0,70){$(f)$}
\end{overpic}
\caption{(color online) (a) The second-order derivative of the
logarithm of the fidelity per lattice site, $\ln
d(\vec{J},\vec{J}')$, with respect to $J_x$ diverges at the critical
point in the thermodynamic limit. However, it remains analytic for
finite-size systems, although more pronounced dips occur with
increasing linear system size. Here $J'_x=0.8$ and
$J_y=J'_y=J_z=J'_z=1/2$. (b) The second-order derivative of the
logarithm of the fidelity per lattice site, $\ln
d(\vec{J},\vec{J}')$, with respect to $J_z$ diverges at the critical
point in the thermodynamic limit. However, it remains analytic for
finite-size systems, although more pronounced dips occur with
increasing linear system size. Here $J'_z=0.8$ and
$J_x=J'_x=J_y=J'_y=1/2$. (c) The dips values scale as $\ln L$ with
increasing linear size $L$ for $J'_x=0.8$ and
$J_y=J'_y=J_z=J'_z=1/2$. (d) The dips values scales as $\ln L$ with
the linear size $L$ for $J'_z=0.8$ and $J_x=J'_x=J_y=J'_y=1/2$. (e)
The positions of the dips approach the critical point $J_{xc} =1$
with increasing linear size $L$. Here $d(\vec{J},\vec{J}')$ is shown
as a function of $J_x$ for $J'_x=0.8$ and $J_y=J'_y=J_z=J'_z=1/2$.
(f) The positions of the dips approach the critical point $J_{zc}
=1$ with increasing linear size $L$. Here $d(\vec{J},\vec{J}')$ is
shown as a function of $J_z$ for $J'_z=0.8$ and
$J_x=J'_x=J_y=J'_y=1/2$.} \label{fig2}
\end{figure}

More precisely, for any fixed $\vec{J}'$, $\ln d(\vec{J};\vec{J}')$
is logarithmically divergent when $\vec{J}'$ is varied such that a
critical point is crossed. Suppose $J_y$ and $J_z$ are fixed, and
only $J_x$ is a control parameter that varies.
 Then we have
\begin{equation}
\frac {\partial^2 {\ln d(\vec{J},\vec{J}')}}{{\partial J_x}^2} = k_1
\ln |J_x -J_{xc}| + {\rm constant} \label{infinite},
\end{equation}
where $k_1$ is a non-universal prefactor that depends on $J_y, J_z$
and $\vec{J}'$, and $J_{xc}$ is the critical value of $J_x$ for
fixed $J_y$ and $J_z$. The numerical results are plotted in
Fig.\;{\ref{fig2}}(a) for $J_y=J_z=1/2$ and $J_{xc}=1$. The least
square fit yields $k_1 \approx  0.02360$. Similarly, we have
presented numerics in Fig.\;{\ref{fig2}}(b) for the second-order
derivative of $\ln d(\vec{J};\vec{J}')$ with respect to $J_z$, with
$J'_z=0.8$ and $J_x=J'_x=J_y=J'_y=1/2$. It turns out that it
diverges logarithmically in the same way as (\ref{infinite}) with
$J_x$ replaced by $J_z$, and $k_1 \approx 0.04726$.

{\it Finite size scaling analysis.}  For a system of finite size $N
\equiv L^2$ (with $L$ the linear size), there is no divergence in
the second-order derivative of $\ln d(\vec{J},\vec{J}')$ with
respect to $J_x$, since QPTs only occur in the thermodynamic limit.
Instead,  as seen in Fig.\;{\ref{fig2}}(a), some pronounced dips
occur at the so-called quasi-critical points $J_{xm}$, with the dips
values logarithmically diverging with increasing linear size $L$,
\begin{equation}
\frac {\partial^2 {\ln d(\vec{J},\vec{J}')}}{{\partial J_x}^2}
\Big|_{J_x =J_{xm}} = k_2 \ln L + {\rm constant} \label{finite},
\end{equation}
where $k_2$ is a non-universal prefactor $k_2$, which takes the
value $k_2 \approx -0.02312$ for $J'_x=0.8$ and
$J_y=J'_y=J_z=J'_z=1/2$ (see Fig.\;{\ref{fig2}}(c)). In addition,
$J_{xm}$ approaches the critical value as $J_{xm} \sim 1-3.96384
L^{-1.06245}$, as follows from Fig.\;{\ref{fig2}}(e). The scaling
ansatz in the system exhibiting logarithmic divergences requires
that the absolute value of the ratio $k_1/k_2$ is the correlation
length critical exponent $\nu$. In this case, $|k_1/k_2|\sim
1.02076$, very close to the exact value $\nu=1$. This is consistent
with the fact that the gap $\Delta$ for the Bogoliubov quasiparticle
scales as $\Delta \sim J_x - J_{xc}$ near the critical point
$J_{xc}$. Similarly, a finite size scaling analysis is performed for
$\ln d(\vec{J},\vec{J}')$ with $J'_z=0.8$ and
$J_x=J'_x=J_y=J'_y=1/2$. In Fig.\;{\ref{fig2}}(c), the least square
fit yields $k_2 \approx -0.04640$. The numerics for ${\partial^2
{\ln d(\vec{J},\vec{J}')}}/{{\partial J_x}^2}|_{J_z =J_{zm}}$ and
$J_{zm}$ are plotted in Figs.\;{\ref{fig2}}(d) and (f),
respectively.

\begin{figure}[ht]
\begin{overpic}[width=42mm,totalheight=32mm]{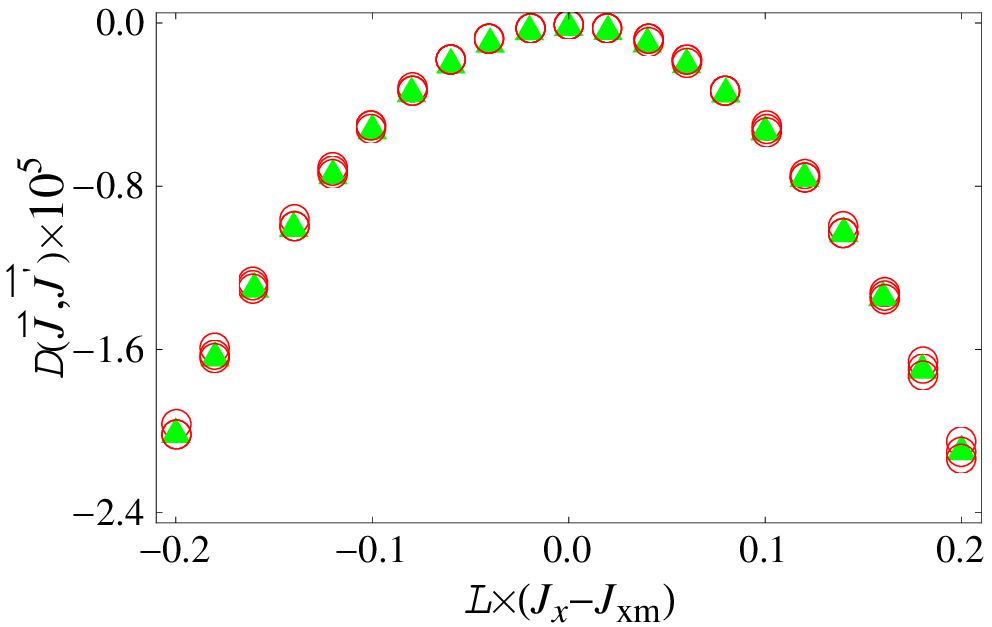}
\put(50,20){$(a)$}
\end{overpic}
\hspace{0in}
\begin{overpic}[width=42mm,totalheight=32mm]{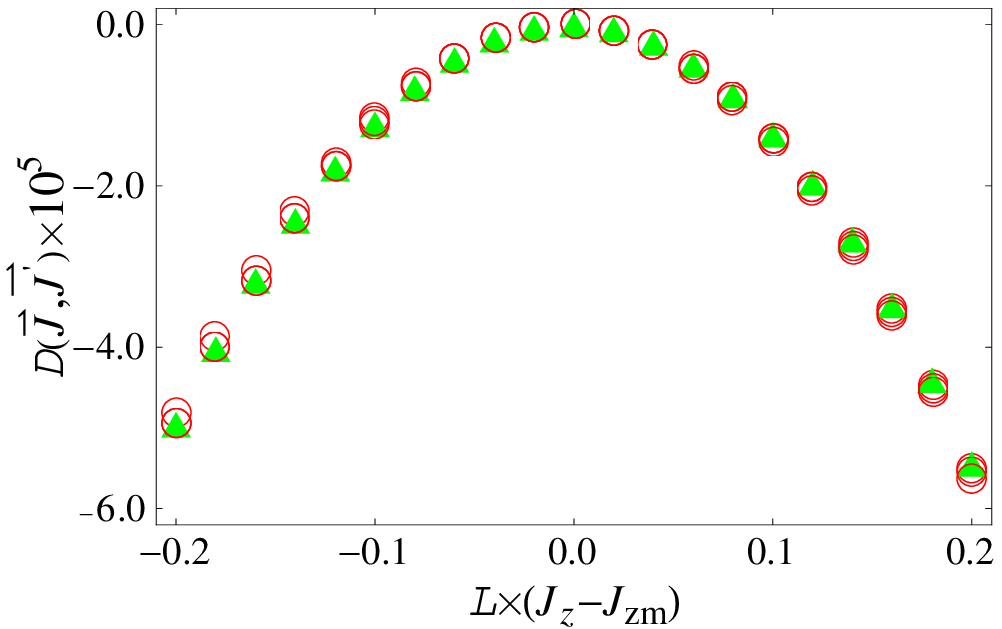}
\put(50,20){$(b)$}
\end{overpic}
\caption{(color online) (a) A finite size scaling analysis is
performed for a quantity defined as $D(\vec{J},\vec{J}') =
1-\exp[\partial^2 _{J_x} \ln d(\vec{J},\vec{J}')-
\partial^2 _{J_x} \ln d(\vec{J},\vec{J}')|_{J_x =
J_{xm}}]$, with $J_y, J_z$ and $\vec{J}'$ fixed.  The scaling ansatz
for logarithmic divergences implies that $D(\vec{J},\vec{J}')$ is a
function of $L (J_x -J_{xm})$ for fixed $J_y, J_z$ and $\vec{J}'$.
(b) A finite size scaling analysis is performed for a quantity
defined as $D(\vec{J},\vec{J}') = 1-\exp[\partial^2 _{J_z} \ln
d(\vec{J},\vec{J}')-
\partial^2 _{J_z} \ln d(\vec{J},\vec{J}')|_{J_z =
J_{xm}}]$, with $J_x, J_y$ and $\vec{J}'$ fixed.  The scaling ansatz
implies that $D(\vec{J},\vec{J}')$ is a function of $L (J_z
-J_{zm})$ for fixed $J_x, J_y$ and $\vec{J}'$.  All the data from
$L=401$ up to $L=1401$ collapse onto a single curve. This shows that
the system at a critical point is scale invariant and that the
correlation length critical exponent $\nu$ is 1. }
  \label{fig3}
\end{figure}

In order to address the scaling ansatz for a system exhibiting
logarithmic divergence~\cite{barber}, we take into account the
distance of the minimum of $\partial^2_{J_x} \ln
d(\vec{J},\vec{J}')$ from the critical point to investigate
 $1-\exp[\partial^2 _{J_x} \ln d(\vec{J},\vec{J}')-
\partial^2 _{J_x} \ln d(\vec{J},\vec{J}')|_{J_x =
J_{xm}}]$ as a function of $L(J_x-J_{xm})$ for different linear
sizes $L$'s. The numerical results for the linear size ranging from
$L=401$ up to $L=1401$ are plotted in Fig.\;{\ref{fig3}}(a). All the
data for different $L$'s collapse onto a single curve, indicating
that the model is scale invariant, i.e., $\xiup/ L = \xiup'/ L'$,
and that the correlation length critical exponent $\nu =1$. The same
conclusion can be drawn from Fig.\;{\ref{fig3}}(b), where the data
collapsing is confirmed for $1-\exp[\partial^2 _{J_z} \ln
d(\vec{J},\vec{J}')-
\partial^2 _{J_z} \ln d(\vec{J},\vec{J}')|_{J_z =
J_{zm}}]$.

{\it Summary.} We have demonstrated that the ground state fidelity
per lattice site is able to detect QPTs in the Kitaev model on the
honeycomb lattice. It is found that, in the thermodynamic limit, the
ground state fidelity per lattice site is non-analytic at a critical
point. More precisely, the second-order derivative of its
logarithmic function with respect to a given control parameter is
logarithmically divergent as the phase boundaries are crossed. A
finite size scaling analysis has also been performed to extract the
correlation length critical exponent from the scaling behaviors of
the fidelity per site. Our exact results offer a benchmark to
numerically investigate QPTs for two-dimensional quantum lattice
systems with topological order in the context of tensor network
representations, which is currently under investigation.

We thank Yupeng Wang for insightful discussions about the Kitaev
model and comments on the manuscript. Support from the Natural
Science Foundation of China is acknowledged.


\begin{thebibliography}{99}
\bibitem{mapping} However, the existence of a mapping from a system with
topological order to a system with symmetry-breaking order is
possible, as shown in Ref.~\cite{xiang} for the Kitaev model.
\bibitem{numerical} It is a difficult task to evaluate the ground state fidelity per site, if a quantum
lattice system is not exactly solvable.  However,  thanks to the
latest advances in classical simulations of quantum lattice
systems~\cite{iTEBD,iPEPS,mera1,mera2},  a practical way to compute
it is now available using the tensor network algorithms for
translationally invariant infinite-size quantum lattice
systems~\cite{zov}.
\bibitem{mera} This amounts to the statement that the fidelity per
lattice site does not depend on which state has been chosen among
degenerate states in the ground state subspace for a system with
topological order. This has been shown for different states of the
toric code~\cite{aguado}. That is, the tensor network representation
for different states shares the same tensors except for their top
tensor, where the relevant topological information is stored. Such a
difference does not contribute to the fidelity per lattice site.
\end{thebibliography}
\end{document}